\documentclass[%
 reprint,
 preprintnumbers,
 amsmath, amssymb,
 aps,
 prl,
 showpacs,
]{revtex4-1}

\usepackage{graphicx}
\usepackage{dcolumn}
\usepackage{bm}
\usepackage{multirow}
\usepackage{threeparttable}
\usepackage{color}
\usepackage{ulem}

\begin{document}


\title{THz Generation and Detection on Dirac Fermions in Topological Insulators}

\author{C. W. Luo$ ^{1}$}
\email{cwluo@mail.nctu.edu.tw}
\author{C. C. Lee$ ^{1} $}
\author{H.-J. Chen$ ^{1} $}
\author{C. M. Tu$ ^{1} $}
\author{S. A. Ku$ ^{1} $}
\author{W. Y. Tzeng$ ^{1} $}
\author{T. T. Yeh$ ^{1} $}
\author{M. C. Chiang$ ^{1} $}
\author{H. J. Wang$ ^{1} $}
\author{W. C. Chu$ ^{1} $}
\author{J.-Y. Lin$ ^{2}$}
\email{ago@nctu.edu.tw}
\author{K. H. Wu$ ^{1} $}
\author{J. Y. Juang$ ^{1} $}
\author{T. Kobayashi$ ^{1,3} $}
\author{C.-M. Cheng$ ^{4} $}
\author{C.-H. Chen$ ^{4} $}
\author{K.-D. Tsuei$ ^{4} $}
\author{H. Berger$ ^{5} $}
\author{R. Sankar$ ^{6} $}
\author{F. C. Chou$ ^{6} $}
\author{H. D. Yang$ ^{7} $}

\affiliation{$ ^1 $Department of Electrophysics, National Chiao Tung University, Hsinchu 300, Taiwan, R.O.C.}
\affiliation{$ ^2 $Institute of Physics, National Chiao Tung University, Hsinchu 300, Taiwan, R.O.C.}
\affiliation{$ ^3 $Advanced Ultrafast Laser Research Center, and Department of Engineering Science, Faculty of Informatics and Engineering, University of Electro-Communications, 1-5-1 Chofugaoka, Chofu, Tokyo 182-8585, Japan}
\affiliation{$ ^4 $National Synchrotron Radiation Research Center, Hsinchu 30076, Taiwan, R.O.C.}
\affiliation{$ ^5 $Institute of Physics of Complex Matter, EPFL, 1015 Lausanne, Switzerland}
\affiliation{$ ^6 $Center for Condensed Matter Sciences, National Taiwan University, Taipei 106, Taiwan, R.O.C.}
\affiliation{$ ^7 $Department of Physics, National Sun Yat-Sen University, Kaohsiung 804, Taiwan, R.O.C.}

\date{\today}

\begin{abstract}
This study shows that a terahertz (THz) wave can be generated from the (001) surface of cleaved Bi$_{\textrm{2}}$Se$_{\textrm{3}}$ and Cu-doped Bi$_{\textrm{2}}$Se$_{\textrm{3}}$ single crystals using 800 nm femtosecond pulses. The generated THz power is strongly dependent on the carrier concentration of the crystals. An examination of the dependence reveals the two-channel free carrier absorption to which Dirac fermions are indispensable. Dirac fermions in Bi$_{\textrm{2}}$Se$_{\textrm{3}}$ are significantly better absorbers of THz radiation than bulk carriers at room temperature. Moreover, the characteristics of THz emission confirm the existence of a recently proposed surface phonon branch that is normalized by Dirac fermions.
\end{abstract}

\maketitle

Three-dimensional topological insulators (TIs) are characterized by a narrow band gap in the bulk and a Dirac cone-like conducting surface state \cite{1,2,3}. The surface state is a new state of quantum matter caused by the strong spin-orbit interaction and protected by time-reversal symmetry. The special properties of TIs have applications in spintronics and quantum computations. Certain TIs with a small band gap are especially useful for terahertz (THz) optoelectronics. One of the key issues about TIs has been the identification of the gapless surface electronic states (Dirac fermions) and the characterizations of their fundamental properties. Angle-resolved photoemission spectroscopy (ARPES) \cite{1,4,5,6} and scanning tunneling microscopy (STM) \cite{7,8,9,10} have successfully confirmed the existence of Dirac fermions in Bi$_{\textrm{1-x}}$Se$_{\textrm{x}}$, Bi$_{\textrm{2}}$Se$_{\textrm{3}}$, and Bi$_{\textrm{2}}$Te$_{\textrm{3}}$. Regarding the transport measurements, a metallic channel associated with the protected surface state has been detected by either controlling the gate voltage in TI devices with a sufficiently low bulk carrier density or by using very thin TI films \cite{11,12,13,14}. However, these experiments used particularly specialized instruments, and their procedures are excessively complex for quick and routine characterizations of Dirac fermions in TIs. Hsieh et al. \cite{15} showed an alternative approach using second harmonic generation (SHG) in arsenic- doped Bi$_{\textrm{2}}$Se$_{\textrm{3}}$ single crystals associated with Dirac fermions, which showed a new venue for examining Dirac fermions by contact-free optical techniques. However, SHG is highly sensitive to the surface quality of samples and doping. THz waves may in principle be an ideal tool for distinguishing Dirac fermions from bulk carriers because they are not sensitive to the surface quality of samples with long wavelengths. Furthermore, THz wave have a photon energy (approximately 4 meV) that is significantly lower than the bulk gap (approximately 300 meV) of TIs; thus, THz radiation would allow specific characterizations within the Dirac cone. Aguilar et al. \cite{16} recently showed the THz responses of Dirac fermions in Bi$_{\textrm{2}}$Se$_{\textrm{3}}$ thin films. However, this type of THz experiments can only be applied to thin TIs of several tens of quintuple layers. This study shows a THz generation from pure Bi$_{\textrm{2}}$Se$_{\textrm{3}}$ and Cu-doped Bi$_{\textrm{2}}$Se$_{\textrm{3}}$ single crystals by pumping with femtosecond laser pulses. Dirac fermions were identified to have an indispensable role on the intensity of THz emission. Free carrier absorption is a crucial mechanism to the optoelectronic devices of TIs and was revealed from the dependence of generated THz power on the carrier concentration \cite{17}. Moreover, the detailed characteristics of THz generation verified a newly proposed phonon branch normalized by Dirac fermions \cite{18}.

\begin{figure}
\begin{center}
\includegraphics[width=8cm]{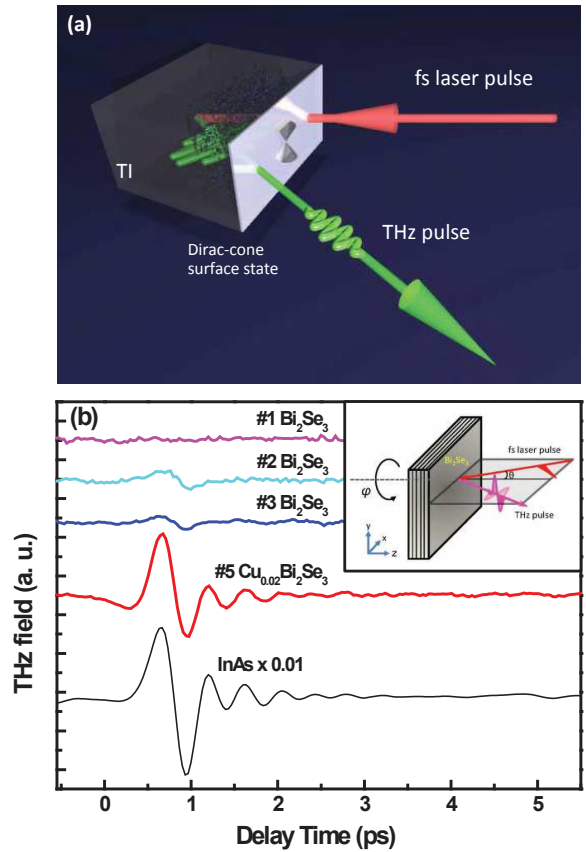}
\caption{\label{fig:fig1}(color online) (a) Schematic illustration of THz generation and FCA inside topological insulators. The dots indicate the free carriers. (b) THz waveform generated from various Bi$_{\textrm{2}}$Se$_{\textrm{3}}$ and Cu$_{\textrm{0.02}}$Bi$_{\textrm{2}}$Se$_{\textrm{3}}$ single crystals, and an InAs wafer. Inset: schematic illustration of THz generation and the detection scheme.}
\end{center}
\end{figure}

Single crystals of pure Bi$_{\textrm{2}}$Se$_{\textrm{3}}$ and Cu-doped Bi$_{\textrm{2}}$Se$_{\textrm{3}}$ were grown using either the Bridgeman, Melt growth, or CVT methods \cite{19}. Single crystals of Cu$_{\textrm{x}}$Bi$_{\textrm{2}}$Se$_{\textrm{3}}$ were obtained using a slow-cooling method from 850 to 650 $^{\circ}$C at a rate of 2 $^{\circ}$C/h and quenching in cold water. Scotch tape was used to cleave the (001) surface of the Bi$_{\textrm{2}}$Se$_{\textrm{3}}$ crystals to ensure a flat and bright surface for optical measurements. The carrier concentrations of the samples listed on Table I were obtained using the Hall measurements. The mobility was measured using the four-probe method. A reflection-type THz generation scheme was used to generate a THz wave on TIs, as shown in Fig. \ref{fig:fig1}(a) and the inset of Fig. \ref{fig:fig1}(b). An 800 nm Ti:sapphire laser (FemtoLasers, Inc.) beam with a repetition rate of 5.2 MHz and a pulse duration of 50 fs was incident at \textit{$\theta$} = 45$^{\circ}$ (to the surface normal) and focused on the surface of the samples with a diameter of 43 $\mu$m. The pumping fluence was tuned by varying the laser output power (the typical value for this study was 0.37 mJ/cm$^{\textrm{2}}$). Following femtosecond pulse pumping, the generated THz wave was collected using a pair of off-axis parabolic mirrors and focused on a 1-mm-thick ZnTe crystal to allow its detection with electro-optical (EO) sampling \cite{20}. The entire generation and detection systems were sealed in a nitrogen-filled plastic box to reduce the humidity to $<$ 6.0\%. All optical measurements were performed at room temperature.
\begin{table}
\begin{center}
\tabcolsep=0.17cm
\caption{Carrier concentration and the THz peak amplitude for samples grown by different methods ($^{\textrm{a}}$ Bridgman, $^{\textrm{b}}$ CVT, $^{\textrm{c}}$ Melt growth). All samples are n-type.}
\begin{tabular}{cccc}
\hline\hline
 &  & Carrier & 
\tabularnewline
Code & Compounds & concentration & THz peak amplitude
\tabularnewline
 &  & (-10$^{\textrm{18}}$ cm$^{\textrm{-3}}$) & (arb. units.)
\tabularnewline
\hline
$^{\textrm{a}}\#1$ & Bi$_{\textrm{2}}$Se$_{\textrm{3}}$ & 75.5$\pm$13.6 & 0.72$\pm$0.62
\tabularnewline
$^{\textrm{b}}\#2$ & Bi$_{\textrm{2}}$Se$_{\textrm{3}}$ & 34.6$\pm$4.37 & 4.49$\pm$0.67
\tabularnewline
$^{\textrm{a}}\#3$ & Bi$_{\textrm{2}}$Se$_{\textrm{3}}$ & 31.0$\pm$1.61 & 3.05$\pm$0.51
\tabularnewline
$^{\textrm{c}}\#4$ & Bi$_{\textrm{2}}$Se$_{\textrm{3}}$ & 15.6$\pm$10.3 & 6.43$\pm$0.43
\tabularnewline
$^{\textrm{a}}\#5$ & Cu$_{\textrm{0.02}}$Bi$_{\textrm{2}}$Se$_{\textrm{3}}$ & 3.66$\pm$0.16 & 31.5$\pm$0.44
\tabularnewline
$^{\textrm{a}}\#6$ & Cu$_{\textrm{0.08}}$Bi$_{\textrm{2}}$Se$_{\textrm{3}}$ & 4.23$\pm$1.17 & 32.3$\pm$0.51
\tabularnewline
$^{\textrm{c}}\#7$ & Cu$_{\textrm{0.1}}$Bi$_{\textrm{2}}$Se$_{\textrm{3}}$ & 1.96$\pm$0.86 & 22.8$\pm$0.17
\tabularnewline
$^{\textrm{c}}\#8$ & Cu$_{\textrm{0.125}}$Bi$_{\textrm{2}}$Se$_{\textrm{3}}$ & 1.17$\pm$0.56 & 30.3$\pm$0.22
\tabularnewline
\hline\hline
\end{tabular}
\end{center}
\end{table}

\begin{figure}
\begin{center}
\includegraphics*[width=7.8cm]{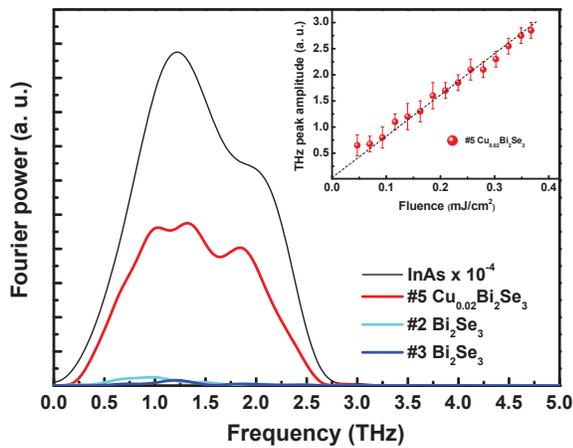}
\caption{\label{fig:fig2}(color online) Fourier power spectra in the frequency domain converted from the time-domain THz signals in Fig. \ref{fig:fig1}(b) using fast Fourier transform for Bi$_{\textrm{2}}$Se$_{\textrm{3}}$ and Cu$_{\textrm{0.02}}$Bi$_{\textrm{2}}$Se$_{\textrm{3}}$ single crystals, and an InAs wafer. Inset: the THz peak amplitude of a Cu$_{\textrm{0.02}}$Bi$_{\textrm{2}}$Se$_{\textrm{3}}$ single crystal as a function of the pumping fluence.}
\end{center}
\end{figure}

Fig. \ref{fig:fig1}(b) shows the typical THz waveform generated from pure Bi$_{\textrm{2}}$Se$_{\textrm{3}}$ and Cu-doped Bi$_{\textrm{2}}$Se$_{\textrm{3}}$ single crystals with a reflection-type setup (THz radiation cannot be detected after TIs). The amplitude of the THz wave generated from pure Bi$_{\textrm{2}}$Se$_{\textrm{3}}$ single crystals is significantly smaller than that from a Cu$_{\textrm{0.02}}$Bi$_{\textrm{2}}$Se$_{\textrm{3}}$ single crystal. Certain Bi$_{\textrm{2}}$Se$_{\textrm{3}}$ crystals such as sample $\#$1 produce nearly zero amplitude (below the S/N ratio in the detection system). The THz generation intensity is strongly dependent on carrier concentration and doping. In general, the THz waveform is composed of a large single pulse and a damped oscillation, which is due to the interference between THz electric fields from different positions of the crystal. These electric fields are caused by the mismatch between the THz phase velocity and the group velocity of the optical pumping pulse \cite{21}. Time-domain THz waveforms (Fig. \ref{fig:fig1}(b)) can be converted to frequency domain spectra (Fig. \ref{fig:fig2}) by using fast Fourier transform (FFT). The central frequency for Cu$_{\textrm{0.02}}$Bi$_{\textrm{2}}$Se$_{\textrm{3}}$ and bandwidth is approximately 1.2 THz and 1.6 THz, respectively. The intensity of pure Bi$_{\textrm{2}}$Se$_{\textrm{3}}$ THz spectra is relatively small, corresponding to small THz signals in the time domain.

\begin{figure}
\begin{center}
\includegraphics*[width=7.8cm]{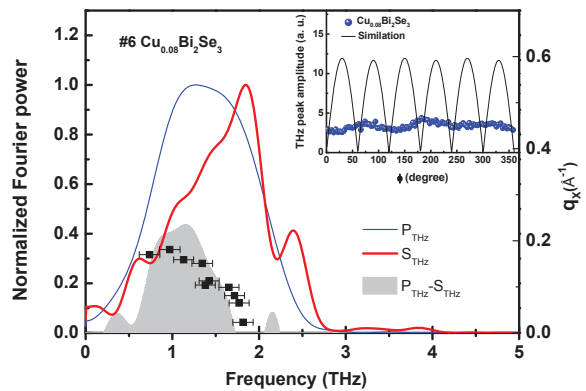}
\caption{\label{fig:fig3}(color online) Polarization-dependent Fourier power spectra of a Cu$_{\textrm{0.08}}$Bi$_{\textrm{2}}$Se$_{\textrm{3}}$ single crystal. P$_{\textrm{THz}}$ (S$_{\textrm{THz}}$): The electric field of the p-(s-)polarized THz radiation is parallel (perpendicular) to the plane of incidence (see the inset of Fig. \ref{fig:fig1}(b)). The solid squares are the surface topological phonon dispersion data taken from Ref. \cite{18}. Inset: THz peak amplitude for a Cu$_{\textrm{0.02}}$Bi$_{\textrm{2}}$Se$_{\textrm{3}}$ single crystal as a function of the azimuth angles $\phi$ along the [001] direction, as shown in the inset of Fig. \ref{fig:fig1}(b).}
\end{center}
\end{figure}

THz signals were measured at various azimuth angles $\varphi$ along the surface normal to understand the THz generation mechanism in TIs (inset of Fig. \ref{fig:fig1}(b)). The data from inset of Fig. \ref{fig:fig3} clearly show that the THz peak amplitude is virtually independent of $\varphi$ and at odds with the optical rectification simulation curve with six-fold symmetry \cite{22}. Therefore, optical rectification is not the dominant mechanism for THz generation in TIs, and the nonlinear effect does not mainly contribute to THz generation in TIs.

The band gap of 0.3 eV in Bi$_{\textrm{2}}$Se$_{\textrm{3}}$ is significantly smaller than the pumping photon energy of 1.55 eV. Free carriers are generated when the femtosecond laser illuminates Bi$_{\textrm{2}}$Se$_{\textrm{3}}$ or Cu-doped Bi$_{\textrm{2}}$Se$_{\textrm{3}}$ crystals. These excited carriers are located inside the bulk within 100 nm \cite{23}. When the electric field is built inside the crystals, the excited carriers in the bulk are driven and form the currents. Two types of built-in electric fields are normally present in semiconductors. The surface depletion field results from the bending of the conduction band on the semiconductor surface as in GaAs \cite{24}. The photo-Dember field is caused by the inhomogeneous distribution of holes and electrons \cite{25}, which is usually present in narrow bandgap semiconductors, such as InAs and InSb with a bandgaps of 0.35 eV and 0.17 eV, respectively. The bandgap is approximately 0.3 eV for Bi$_{\textrm{2}}$Se$_{\textrm{3}}$ single crystals, which is close to that of InAs (reference sample in Fig. \ref{fig:fig1}). Additionally, the intrinsic charge inhomogeneity in the vicinity of the surface and, as a result, band-bending effects were reported [26]. Consequently, both photo-Dember and surface depletion effects are possible mechanism for THz generation in Bi$_{\textrm{2}}$Se$_{\textrm{3}}$ and Cu-doped Bi$_{\textrm{2}}$Se$_{\textrm{3}}$ single crystals. The currents formed by the excited carriers are suppressed within several picoseconds, due to carrier scattering with impurities (e.g., Se vacancies) or with the layer boundary. The transient current further generates THz radiation  by $\textit{E}_{\textrm{\textit{THz}}}(t)\propto\partial\textit{J}(t)/\partial\textit{t}$. Higher pumping fluences generate more free carriers, leading to larger changes of the transient current; thus , a stronger THz radiation should be generated. The THz peak amplitude increases linearly with the pumping fluences (inset of Fig. \ref{fig:fig2}) \cite{27}.

The excited carriers in the bulk can diffuse either along the [001] direction or on the (001) plane to form two types of currents. In principle, the diffusion along the [001] direction is suppressed by the layer boundary or impurity scattering to generate p-polarized THz radiation, and the diffusion on the (001) plane is suppressed mainly by impurity scattering to generate s-polarized THz radiation. The addition of a single wire-grid polarizer between the sample and an EO detection system allows the p- and s-polarized THz radiation from TIs (e.g., from the Cu$_{\textrm{0.08}}$Bi$_{\textrm{2}}$Se$_{\textrm{3}}$ single crystal) to be distinguished (Fig. \ref{fig:fig3}). Intriguingly, the difference between p-polarized and s-polarized THz radiation is not only in the central frequency, but also in the shape of the spectra. Subtraction of the s-polarized THz spectrum from the p-polarized THz spectrum results in the gray area between 0.53 THz and 1.72 THz (Fig. \ref{fig:fig3}), indicating additional absorption for the s-polarized THz wave. Zhu et al. \cite{18} recently measured the surface phonon dispersion in Bi$_{\textrm{2}}$Se$_{\textrm{3}}$ crystals using the coherent helium beam surface scattering technique. They discovered a low-energy isotropic convex-dispersive surface phonon branch normalized by Dirac fermions and with a frequency range from 1.8 to 0.74 THz. This frequency range is consistent with that of the gray area in Fig. \ref{fig:fig3}. Therefore, the missing power of an electromagnetic wave with an electrical field parallel to the surface manifests the additional effect due to the Dirac-fermion-normalized surface phonons.

\begin{figure}
\begin{center}
\includegraphics*[width=8cm]{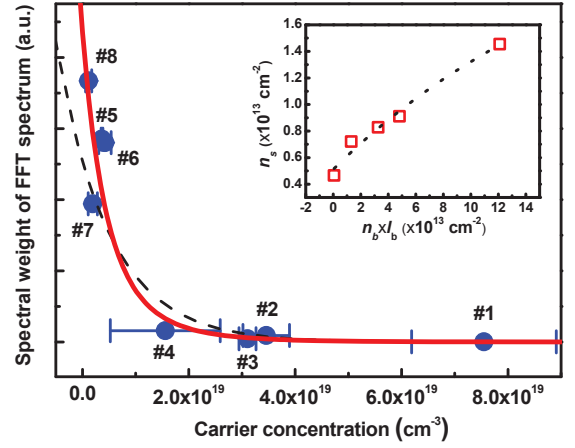}
\caption{\label{fig:fig4}(color online) Spectral weight of the fast Fourier transform (FFT) spectra vs. the carrier concentration. The black dashed line represents the fit by Eq. (\ref{eq1}). The red solid line represents the fit by Eq. (\ref{eq2}). Inset: the carrier concentration \textit{n}$_{\textit{s}}$ of the Dirac fermions vs. \textit{n}$_{\textit{b}}$ of the bulk carriers. The dotted line represents the empirical fit of \textit{n}$_{\textit{s}}\times$10$^{\textrm{-13}}$=0.51+2.10[1-exp(-\textit{n}$_{\textit{b}}\times$\textit{l}$_{\textit{b}}\times$10$^{\textrm{-13}}$/20.48)].}
\end{center}
\end{figure}

Table I shows that the carrier concentration decreases by more than one order of magnitude when Cu is doped into Bi$_{\textrm{2}}$Se$_{\textrm{3}}$ crystals (n-type, caused by Se vacancies) because Cu replaces Bi in the lattice \cite{28,29,30}. The carrier concentrations from 15.6$\pm$10.3$\times$10$^{\textrm{18}}$ cm$^{\textrm{-3}}$ to 75.5$\pm$13.6$\times$10$^{\textrm{18}}$ cm$^{\textrm{-3}}$ were observed for the pure Bi$_{\textrm{2}}$Se$_{\textrm{3}}$ crystals, potentially because of the different growing conditions and methods. The THz signals from pure Bi$_{\textrm{2}}$Se$_{\textrm{3}}$ crystals are generally relatively small. Conversely, the Cu$_{\textrm{0.02}}$Bi$_{\textrm{2}}$Se$_{\textrm{3}}$ crystal with a lower carrier concentration produces stronger THz emission than pure Bi$_{\textrm{2}}$Se$_{\textrm{3}}$ crystals. The spectral weight of the FFT spectra in Fig. \ref{fig:fig2} was plotted as a function of the carrier concentration to further quantify these results. Fig. \ref{fig:fig4} clearly shows that the THz output power (i.e., the spectral weight of the FFT spectrum) increases with the decreasing carrier concentration.

The THz wave generated inside the bulk likely suffers free carrier absorption (FCA) during propagation. This effect can generally be reduced by suppressing the carrier concentration to increase the output intensity of the THz, as demonstrated in III-V semiconductors such as InAs \cite{31}. The dependence of the THz intensity on the carrier concentration can be phenomenologically described using the Beer-Lambert Law:
\begin{eqnarray}
\label{eq1}
\textit{I}_{\textrm{THz}} = \textit{I}_{\textrm{0,THz}}e^{-\sigma\textit{l}\textit{n}}
\end{eqnarray}
where \textit{I}$_{\textrm{THz}}$ and \textit{I}$_{\textrm{0,THz}}$ are the transmitted THz intensities outside and inside the samples, respectively, as shown in Fig. \ref{fig:fig1}(a), \textit{l} is the path length, \textit{n} is the carrier concentration, and $\sigma$ is the absorption cross-section. The experimental data in Fig. \ref{fig:fig4} can be fitted as the black dashed line by Eq. (\ref{eq1}), and is qualitatively similar to the case of InAs \cite{31}. However, a closer inspection reveals that the fit by Eq. (\ref{eq1}) leads to a larger deviation from the data in the low concentration regime. Consequently, the experimental data in Fig. \ref{fig:fig4} cannot be explained solely by FCA due to the bulk carriers.

It is known that Dirac fermions exist in Bi$_{\textrm{2}}$Se$_{\textrm{3}}$ crystals \cite{1}. The Dirac band contribution to FCA on the THz emission must be considered. Fig. \ref{fig:fig1}(a) shows that the THz wave can be absorbed by the electrons on the Dirac cone. The effective two-channel FCA including the contribution from both the bulk carriers and Dirac fermions is written as a modified Beer-Lambert equation:
\begin{eqnarray}
\label{eq2}
\textit{I}_{\textrm{THz}} = \textit{I}_{\textrm{0,THz}}e^{-(\sigma_{\textit{b}}\textit{l}_{\textit{b}}\textit{n}_{\textit{b}}+\sigma_{\textrm{s}}\textit{l}_{\textrm{s}}\frac{\textit{n}_{\textit{s}}}{\textit{l}_{\textit{s}}})}
\end{eqnarray}
where $\sigma_{\textit{b}}$ is the absorption cross-section of bulk, \textit{l}$_{\textit{b}}$ is the path length of the bulk, \textit{n}$_{\textit{b}}$ is the carrier concentration in bulk, $\sigma_{\textit{s}}$ is the absorption cross-section of the surface state, \textit{l}$_{\textit{s}}$ is the path length of the surface state, and \textit{n}$_{\textit{s}}$ is the carrier concentration in the surface state. The empirical relation between \textit{n}$_{\textit{b}}$ and \textit{n}$_{\textit{s}}$ can be described by \textit{n}$_{\textit{s}}\times$10$^{\textrm{-13}}$=0.51+2.10[1-exp(-\textit{n}$_{\textit{b}}\times$\textit{l}$_{\textit{b}}\times$10$^{\textrm{-13}}$/20.48)] as shown in the inset of Fig. \ref{fig:fig4} \cite{32}. The observed spectral weight of THz in Fig. \ref{fig:fig4} is significantly better fit by Eq. (\ref{eq2}) (the red solid line) with \textit{l}$_{\textit{s}}$ = 2 nm (thickness of the surface state) \cite{33} and \textit{l}$_{\textit{b}}$ = 23.5 nm (THz emitted from the bulk within 23.5 nm; i.e., the penetration depth of 800 nm pumping light) \cite{34}. The acquired fitting parameters are $\sigma_{\textit{b}}$ =2.59$\times$10$^{\textrm{-14}}$ cm$^{\textrm{2}}$ and $\sigma_{\textit{s}}$ =1.46$\times$10$^{\textrm{-13}}$ cm$^{\textrm{2}}$, and the ratio of $\sigma_{\textit{s}}$/$\sigma_{\textit{b}}$ = 5.64. Therefore, this study concludes that FCA caused by Dirac fermions is more efficient than that caused by bulk carriers (at least at room temperature).

FCA can be explained in the context of the Drude model. The FCA cross-section
\begin{eqnarray}
\label{eq3}
\sigma = \frac{e\mu}{nc\varepsilon_{\textrm{0}}(1+\omega^{\textrm{2}}\tau^{\textrm{2}})}
\end{eqnarray}
where \textit{e} is the electron charge, $\mu$ is the mobility, $\omega$ is the angular frequency of the THz wave, $\tau$ is the scattering time of the carriers, \textit{n} is the refractive index of the material, \textit{c} is the seed of light, and $\varepsilon_{\textrm{0}}$ is the permittivity constant. Equation (\ref{eq3}) indicates that the observed cross-sectional ratio is as follows:
\begin{eqnarray}
\label{eq4}
\frac{\sigma_{\textit{s}}}{\sigma_{\textit{b}}} = \frac{\mu_{\textit{s}}}{\mu_{\textit{b}}} \frac{(1+\omega^{\textrm{2}}\tau_{\textit{b}}^{\textrm{2}})}{(1+\omega^{\textrm{2}}\tau_{\textit{s}}^{\textrm{2}})}
\end{eqnarray}
where \textit{s} denotes the physical properties of the Dirac fermions of the surface state and \textit{b} denotes the physical properties of the bulk carriers. In this study, $\mu_{\textit{b}}\approx$ 1500 cm$^\textrm{2}$/V s at room temperature for the typical Bi$_{\textrm{2}}$Se$_{\textrm{3}}$ single crystals. The mobility can be expressed as $\mu_{\textit{b}}$ = $e\tau$/$m$, where \textit{m} is the effective mass of the carriers. With the effective mass of the bulk carrier $m_{\textit{b}}$=0.14$m_{\textit{0}}$ \cite{35,36}, where $m_{\textit{0}}$ is the electron bare mass, $\tau_{\textit{b}}\approx$0.12 ps at room temperature. According to Ref. 36 and the references therein, $\tau_{\textit{s}}\ll\tau_{\textit{b}}$ at low temperatures (low \textit{T}); so is it assumed at room temperature. The above analysis results in $\omega \tau_{\textit{b}}\approx$ 1.2 and $\omega \tau_{\textit{s}}\ll$  1. Usually, $\mu_{\textit{s}}$ is smaller than $\mu_{\textit{b}}$ at low \textit{T} in Bi$_{\textrm{2}}$Se$_{\textrm{3}}$ single crystals of approximately 100 $\mu$m in thickness \cite{11,35,36}. However, the experimental data for $\mu_{\textit{s}}$ at room temperature have been relatively rare. Equation (\ref{eq4}) indicates that the ratio $\sigma_{\textit{s}}$/$\sigma_{\textit{b}}$ = 5.64 may lead to comparable values for $\mu_{\textit{s}}$ and $\mu_{\textit{b}}$ at room temperature. Generally, the effective mass varies weakly with \textit{T}. Conversely, $\tau_{\textit{b}}$ decreases rapidly with increasing \textit{T} because of electron-phonon scattering. The scattering of Dirac fermions may be more dominated by the impurity scattering than that of the bulk carriers; thus, $\tau_{\textit{s}}$ is less temperature dependent and does not decrease as rapidly with increasing \textit{T} compared to $\tau_{\textit{b}}$. Therefore, the values of $\mu_{\textit{s}}$ and $\mu_{\textit{b}}$ may be comparable at room temperature. These results strongly suggest that a THz wave generated inside the Bi$_{\textrm{2}}$Se$_{\textrm{3}}$ crystals can be used not only to identify but also to effectively examine fundamental properties of Dirac fermions.

In summary, THz radiation can be generated from Bi$_{\textrm{2}}$Se$_{\textrm{3}}$ and Cu-doped Bi$_{\textrm{2}}$Se$_{\textrm{3}}$ single crystals. Dirac fermions of the surface state are indispensable to explaining the strong dependence of the THz emission power on the carrier concentration. Furthermore, the detailed characteristics of the THz emission confirmed the presence of a Dirac-fermion-normalized phonon branch and valuable information regarding the fundamental properties of Dirac fermions.

This work was support by the National Science Council of Taiwan, under grant: Nos. NSC101-2112-M-009-016-MY2, NSC101-2112-M-009-017-MY2 and NSC 100-2112-M110-004-MY3, and by the MOEATU program at NCTU of Taiwan, R.O.C. Technical help from C. K. Wen and discussions with H. T. Jeng and T. M. Uen are appreciated.


\begin{thebibliography}{999}
\bibitem{34}The 800 nm penetration depth was determined by the incident angle-dependent reflectivity measurements (with s- and p-polarizations) and the fitting of complex Fresnel equation.
\end{thebibliography}
\end{document}